\documentclass[twocolumn]{jpsj3}
\usepackage{txfonts}
\usepackage{bm}
\usepackage{color}

\def\up{\uparrow}
\def\down{\downarrow }
\def\Vec#1{\bm{#1}}

\def\sla#1{\rlap/#1}

\title{Spin-polarized Majorana Bound States inside a Vortex Core in Topological Superconductors}

\author{Yuki \surname{Nagai}, Hiroki \surname{Nakamura}, and Masahiko \surname{Machida}}
\inst{
\address{CCSE, Japan  Atomic Energy Agency, 5-1-5 Kashiwanoha, Kashiwa, Chiba 277-8587, Japan} 
}

\abst{
We reveal that Majorana bound states inside the vortex core in an odd-parity topological superconductivity 
classified as ``pseudo-scalar" type in the gap function are distinctly spin-polarized by solving
the massive Dirac Bogoliubov-de Gennes (BdG) equation considering the spin-orbit coupling. 
This result is universal for ``Dirac superconductivity" whose 
rotational degree of freedom is characterized by the total angular momentum ${\bm J} = {\bm S} + {\bm L}$ and in marked contrast to 
the spin-degeneracy of the core bound states as the consequence of the conventional BdG equation. 
The spin-polarized vortex core can be easily detected by spin-sensitive probes such as the neutron scattering 
and other measurements well above the first critical magnetic field $H_{c1}$.
}

\begin{document}
\maketitle
\section{Introduction}
The discovery of topological superconductors opened a new research avenue on superconducting states.
The topologically-protected nature together with $U$(1) broken symmetry results in gapless zero-energy quasi-particles 
identified as Majorana fermions 
at surface edges, while the superconducting gap opens in the bulk body.
The emerged Majorana fermion is a counterintuitive particle whose annihilation and creation operators are identical.  
Such a unique particle has a promising role in topological quantum computing utilizing its non-Abelian statistics.\cite{Teo}
This fascinating feature has highly stimulated many theorists and experimentalists to 
intensively study the topological 
superconductivity\cite{Bernevig15122006,PhysRevLett.105.266401,PhysRevB.76.045302,PhysRevLett.98.106803,RevModPhys.82.3045,PhysRevLett.95.146802,Konig02112007,PhysRevLett.105.146801,PhysRevB.75.121306,PhysRevB.81.041309,PhysRevLett.105.136802}.
However, its research history is not so long as the topological insulator, and most of rich physics still 
remain elusive.

Very recently, experimental works on topological insulators, Bi$_{2}$Se$_{3}$ and SnTe have revealed that they 
turn into superconductors with carrier doping. 
Their superconducting gap functions are not conventional 
since zero-bias conductance peaks (ZBCP's) have been detected 
by the point contact spectroscopy\cite{PhysRevLett.109.217004,PhysRevLett.107.217001}.
The ZBCP is known to be observed in not only unconventional non-$s$-wave superconductors but 
also topological superconductors\cite{PhysRevLett.105.097001,PhysRevLett.104.057001,PhysRevB.86.064517,
Wray2010,PhysRevLett.107.217001,PhysRevLett.109.217004,Hsieh2012,PhysRevLett.106.127004}.
The latter typical example is the chiral $p$-wave topological superconductor such as Sr$_{2}$RuO$_{4}$, in which 
gapless quasi-particles assigned as Majorana fermions induce ZBCP's\cite{JPSJSuppl.89.127,PhysRevB.79.224508,PhysRevLett.74.3451}.
Accordingly, ZBCP's observed in Cu$_{x}$Bi$_{2}$Se$_{3}$ ($T_{c} \sim 3$K) and 
Sn$_{1-x}$In$_{x}$Te ($T_{c} \sim 1.2$K) can be also regarded to be originated from 
their non-trivial topology.

In topological superconductors, the Majorana fermion appears at not only surface edges but also vortex cores. 
Its emergence inside the vortex core has been 
numerically confirmed by using the Bogoliubov-de Gennes (BdG) formalism for $p$-wave triplet 
superconductors\cite{PhysRevB.65.014504,PhysRevB.65.014508}.
Since the vortex is a movable object, the Majorana fermion confined inside the core is also mobile.
The character is hopeful for the topological quantum computing, because vortex manipulation techniques have been 
rapidly developed in the last decade. In this paper, we present a new insight on the Majorana bound fermion 
inside the vortex core. We reveal 
that the Majorana state is spin-polarized in topological superconductors originated 
from strong spin-orbit coupling. Such a case is not particular but rather universal, since the spin degeneracy 
supposed to be kept in the conventional BdG formalism is generally broken 
when spin and orbital angular momenta are coupled. 

\begin{figure}
\begin{center}
     \begin{tabular}{p{0.75 \columnwidth} }
      \resizebox{0.75 \columnwidth}{!}{\includegraphics{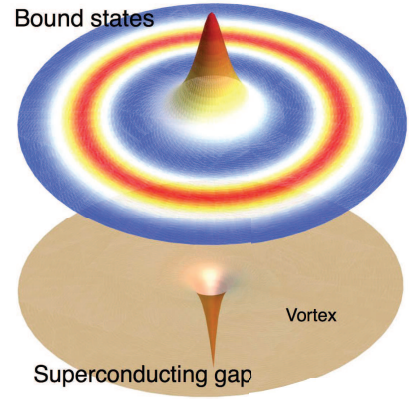}} 
    \end{tabular}
\end{center}
\caption{\label{fig:fig1}(Color online) Spin-polarized Majorana bound states. The red (blue) region denotes 
dominance of up-spin (down-spin) component.}
\end{figure}

Two orbital degrees of freedom in addition to two spin ones are required in a minimum model of the topological superconductivity under the strong spin-orbit coupling\cite{PhysRevLett.98.106803,PhysRevB.83.134516,PhysRevLett.107.217001}. 
Then, the starting BdG Hamiltonian corresponds to a massive Dirac-type one including 
off-diagonal gap functions, in which the Lorentz transformation 
invariance together with anti-commutation between two fermions gives mathematical restrictions on possible gap functions.
As described in Ref.~\citen{PhysRevB.86.094507}, the 
allowed gap functions are characterized 
as a scalar, a pseudo-scalar, and polar-vectors,  when considering the on-site Cooper pairing. 
The polar-vector case always produces a specific direction
around which the rotational isotropy is broken below the superconducting transition. 
Consequently, the restrictions allow six kinds of pairings, which are classified into two even-parity pairings being not topological and 
four odd parity ones being topological. 
In Cu$_{x}$Bi$_{2}$Se$_{3}$, Sasaki {\it et al}. theoretically examined  
which type among these pairings successfully reproduces observed ZBCP's and 
suggested that the odd-parity spin-triplet pairing is the most likely. 
Another superconductor, Sn$_{1-x}$In$_{x}$Te can be also 
checked by the same scheme\cite{PhysRevLett.109.217004}.
In this paper, we study vortex core bound-states in an odd-parity superconductor whose gap function 
is characterized as a ``pseudo-scalar" keeping the rotational isotropy in all directions. 
In this case, the zero-energy Majorana bound states 
emerge inside the vortex cores as well as surface edges, and 
the core-bound Majorana fermion is distinctly spin-polarized around the vortex line. 
In other topological cases such as polar vectors, 
the spin-polarization becomes obscure since the rotational isotropy is broken around the vortex line and the polarized core states are mixed with each other. 
The Majorana fermions are also spin-polarized in topological wires\cite{PhysRevLett.108.096802}. 
Consequently, we address that the spin-polarized  character of the Majorana bound states is universal. 
We analytically and numerically demonstrate the spin-polarized vortex core and discuss 
the theoretical mechanism as well as its experimental detection possibilities.  

\section{Model}
An effective theory for the topological superconductor with the intrinsic spin-orbit coupling
is given by the massive Dirac type of BdG Hamiltonian together with a 
Nambu representation for the two orbital degrees of freedom as
\begin{align}
H &= \int d{\bm r}
\left(\begin{array}{cc}\bar{\psi}({\bm r}) & \bar{\psi}_{\rm c}({\bm r})\end{array}\right)
\left(\begin{array}{cc}\hat{H}^{-}({\bm r})  & \Delta^{-}({\bm r}) \\
\Delta^{+}({\bm r})& \hat{H}^{+}({\bm r}) \end{array}\right)
\left(\begin{array}{c}\psi({\bm r}) \\
\psi_{\rm c}({\bm r})
\end{array}\right), \label{eq:hami}
\end{align}
where
\begin{align}
\hat{H}^{\pm}({\bm r}) &= M(-i {\bm \nabla})  + \sum_{\nu=1}^{3}P_{\nu}(-i {\bm \nabla})\gamma^{\nu}
\pm P_{0}(-i {\bm \nabla}) \gamma^{0} .\label{eq:dirac}
\end{align}
Here, $\gamma^{i}$ is a $4 \times 4$ Dirac gamma matrix which can be described as $\gamma^{0} = \hat{\sigma}_{z} \otimes 1$,
$\gamma^{i = 1,2,3} = i \hat{\sigma}_{y} \otimes \hat{s}_{i}$, and $\gamma^{5} = \hat{\sigma}_{x} \otimes 1$ 
with $2 \times 2$ Pauli matrices $\hat{\sigma}_{i}$ in the orbital space and $\hat{s}_{i}$ in the spin space, 
$\psi(\bm r)$ is the Dirac spinor, $\bar{\psi}(\Vec{r}) \equiv \psi^{\dagger}(\Vec{r})\gamma^{0}$, $\bar{\psi}_{c}(\Vec{r}) 
\equiv \psi_{c}^{\dagger} \gamma^{0}$, and $\psi_{c} \equiv {\cal C} \bar{\psi}^{T}$,
where ${\cal C} (\equiv i \gamma^{2} \gamma^{0})$ is the representative matrix of the charge conjugation. 
$M$ and $P_{\nu}$ are functions whose forms depend on the materials. 
$\Delta^{-}$ is a gap function, and $\Delta^{+} \equiv \gamma^{0} (\Delta^{-})^{\dagger} \gamma^{0}$.
Considering only the on-site pairing interaction, the possible gap form is reduced to six types of functions as seen in Table I of Ref.~\citen{PhysRevB.86.094507}. 
These gap functions are classified into a pseudo-scalar, a scalar, and a polar vector (four-vector)
associated with the Lorentz transformation,
\begin{align}
\Delta^{-} &= \Delta_{0}, \: \Delta_{0} \gamma^{5}, \: \Delta_{0}\sla{\alpha} \gamma^{5}, 
\end{align}
where $\Delta_{0}$ is a scalar,
the Feynman slash $\sla{\alpha}$ is defined by $\sum_{\mu} \gamma^{\mu }\alpha_{\mu}$, and 
the gap function including $\sla{\alpha}$ is characterized as a unit four-vector $\alpha_\mu$ (See, Table I). 
Since the vector type of the gap form represented by $\alpha_\mu$ with finite $\alpha_i$-components ($i=1,2,3$) points to 
a specific direction, the rotational isotropy is broken except for the rotation around the specific direction, resulting in the anisotropic quasi-particle spectrum below its superconducting transitions, even if the normal state is isotropic. 
This superconductivity induced anisotropy yields the angle-dependent transport conductivity.
For example, the anisotropic thermal conductivity is a clear evidence of the vector type $\alpha_\mu$ \cite{PhysRevB.86.094507}.  
From the Hamiltonian Eq.~(\ref{eq:hami}), the correspondent BdG equations are given as
\begin{align}
\left(\begin{array}{cc} \gamma^{0} \hat{H}^{-}({\bm r})  &\gamma^{0} \Delta^{-}({\bm r})\\
\gamma^{0} \Delta^{+}({\bm r}) & \gamma^{0} \hat{H}^{+}({\bm r}) \end{array}\right)
\left(\begin{array}{c}
u({\bm r}) \\
u_{\rm c}({\bm r})
\end{array}\right)
&= E \left(\begin{array}{c}
u({\bm r}) \\
u_{\rm c}({\bm r})
\end{array}\right), \label{eq:bdgdirac}
\end{align}
where we note that $v$ in the conventional eigen-state form, $(u, v)^T$ is given as $v \equiv i \gamma^{2} u_{c}$.  
\begin{table*}[t]
\caption{
Correspondence between our BdG gap functions $\hat{\Delta}^{-}$ and another representations. 
``P-scalar'' denotes a pseudoscalar whose parity is 
odd and ``$i$-polar'' denotes a polar vector pointing to the $i$-direction in four-dimensional space.}
\label{table:1}
\begin{center}
\begin{tabular}{lccccccl}
\hline
&$\hat{\Delta}^{-}$  &Parity & Fu-Berg\cite{PhysRevLett.105.097001} &  Sasaki {\it et al.}\cite{PhysRevLett.107.217001}& energy gap\\
\hline
Scalar & $\gamma^{5}$ &$+$ &$A_{1g}$ & $\Delta_{1a}$: $\Delta_{\up \down}^{11}= -\Delta_{\down \up}^{11}= \Delta_{\up \down}^{22} 
= -\Delta_{\down \up}^{22}$ & full gap\\
$t$-polar & $\gamma^{0} \gamma^{5}$ &$+$ & $A_{1g}$ 
& $\Delta_{1b}$: $\Delta_{\up \down}^{11}= -\Delta_{\down \up}^{11}= -\Delta_{\up \down}^{22} 
= \Delta_{\down \up}^{22}$ &full gap
 \\
P-scalar & $1$ & $-$ & $A_{1u}$ & 
$\Delta_{2}$: $ \Delta_{\up \down}^{12}= -\Delta_{\down \up}^{12}= \Delta_{\up \down}^{21} 
= -\Delta_{\down \up}^{21}$ &full gap
\\
$x$-polar & $\gamma^{1}\gamma^{5}$& $-$ 
&$E_{u}$ & 
$\Delta_{4b}$: $\Delta_{\up \up}^{12}= -\Delta_{\down \down}^{12}= - \Delta_{\up \up}^{21} 
= \Delta_{\down \down}^{21}$ & point-node
\\
$y$-polar  & $\gamma^{2}\gamma^{5}$ & $-$ 
&$E_{u}$ & 
$\Delta_{4a}$: $\Delta_{\up \up}^{12}= \Delta_{\down \down}^{12}= - \Delta_{\up \up}^{21} 
= -\Delta_{\down \down}^{21}$ & point node
\\
$z$-polar & $\gamma^{3}\gamma^{5}$ & $-$
&$A_{2u}$ & 
$\Delta_{3}$: $\Delta_{\up \down}^{12}= \Delta_{\down \up}^{12}= -\Delta_{\up \down}^{21} 
= -\Delta_{\down \up}^{21}$ &point node
\\
\hline 
\end{tabular} 
\end{center}
\end{table*}

\section{Analytical results}
Now, let us examine the vortex-core bound states. 
%
%
In order to analytically concentrate on the low-energy physics, 
we set the functions as $M(- i \Vec{\nabla}) = M_{0}$, $P_{1,2,3}(- i \Vec{\nabla})  = - i \bar{P}_{1,2,3} \partial_{x,y,z}$, and $P_{0}(- i \Vec{\nabla}) = \mu$ (the chemical potential), where $M_{0}$, $\bar{P}$,  and $\mu$ are constants. 
Then, the Dirac Hamiltonian $H^{\pm}(\Vec{r})$ is linearized as
\begin{align}
\hat{H}_{\rm eff}^{\pm}({\bm r}) &= M_{0}  - i  \partial_{x} \gamma^{1} -i \partial_{y} \gamma^{2}
- i \partial_{z} \gamma^{3} \pm \mu \gamma^{0} ,
\end{align}
with the rescaled axes $(x,y,z) \rightarrow (\bar{P}_{1} x, \bar{P}_{2}y,\bar{P}_{3} z)$.
Hereafter, we do not self-consistently solve the gap equation but just use a well-known analytical form $f(r)=r/\sqrt{r^{2}+1}$ for the radial profile of the gap function given as $\Delta^{-}(r) \equiv \bar{\Delta}^{-} f(r) e^{i \theta}$, where $\theta$ denotes the polar angle around the vortex line.

At the zero energy, there is a relation expressed as $u_{c}(\Vec{r}) = i \gamma^{2} u^{\ast}(\Vec{r})$, and 
 the solution of Eq.~(\ref{eq:bdgdirac}) is given as 
$u(\Vec{r}) = \zeta(r) u^{\rm N}(\Vec{r})$,
where
$u^{\rm N}(r,\theta,z)$ is that of the normal state and  $\zeta(r)$ is a scalar real function.
In the case of the pseudo-scalar type of gap function $\bar{\Delta}^{-} = 1$ (so-called $\hat{\Delta}_{2}$, inter-orbital spin-singlet gap function shown as 
\(
\Delta^{12}_{\uparrow\downarrow}
=
- \Delta^{12}_{\downarrow\uparrow}
\), 
\(
\Delta^{21}_{\uparrow\downarrow}
=
- \Delta^{21}_{\downarrow\uparrow}
\), 
\(
\Delta^{12}_{\uparrow\downarrow}
=
\Delta^{21}_{\uparrow\downarrow}
\) 
 in Ref.~\citen{PhysRevLett.109.217004}),
there are two bound-state solutions at the zero-energy expressed as 
\begin{align}
u_{\up}({\bm r}) &= \frac{e^{-K(r)}}{\sqrt{\lambda_{+}}}
\left(\begin{array}{c}
\sqrt{\mu + M_{0}} J_{0}(\bar{r})\\
0 \\
0\\
i e^{i \theta}\sqrt{\mu - M_{0}} J_{1}(\bar{r})
\end{array}\right), \label{eq:up} \\
u_{\down}({\bm r}) &= \frac{e^{-K(r)}}{\sqrt{\lambda_{-}}}
\left(\begin{array}{c}
0\\
e^{i \theta}\sqrt{\mu + M_{0}} J_{1}(\bar{r}) \\
-i \sqrt{\mu - M_{0}} J_{0}(\bar{r})\\
0
\end{array}\right), 
\label{eq:down}
\end{align}
where $\bar{r} \equiv r \sqrt{\mu^{2} - M_{0}^{2}} $, $K(r) \equiv \int_{0}^{r} |f(r')|  dr'$, and $J_{n}(r)$ is the Bessel function of the first kind (see the Appendix A).
Here, $\lambda_{\pm}$ is determined by 
\begin{align}
\lambda_{\pm} &= 4 \pi \int_{0}^{\infty} dr r e^{-2 K(r)} 
 \left[
(\mu \pm M_{0}) J_{0}(\bar{r})^{2} +(\mu \mp M_{0}) J_{1}(\bar{r})^{2} 
\right]. 
\end{align}

\begin{figure}
\begin{center}
     \begin{tabular}{p{0.5 \columnwidth} p{0.5 \columnwidth}}
      \resizebox{0.45 \columnwidth}{!}{\includegraphics{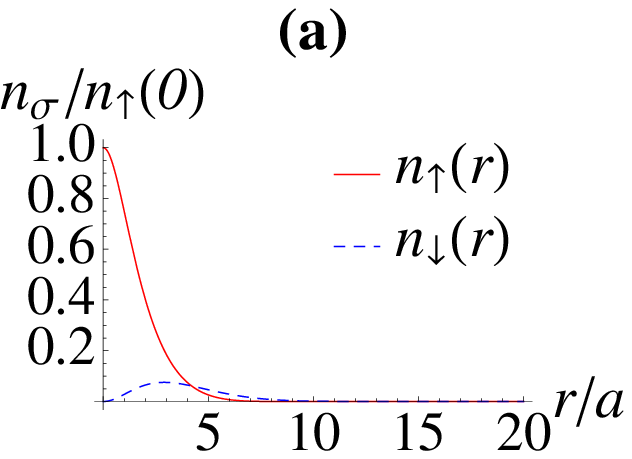}} &
      \resizebox{0.45 \columnwidth}{!}{\includegraphics{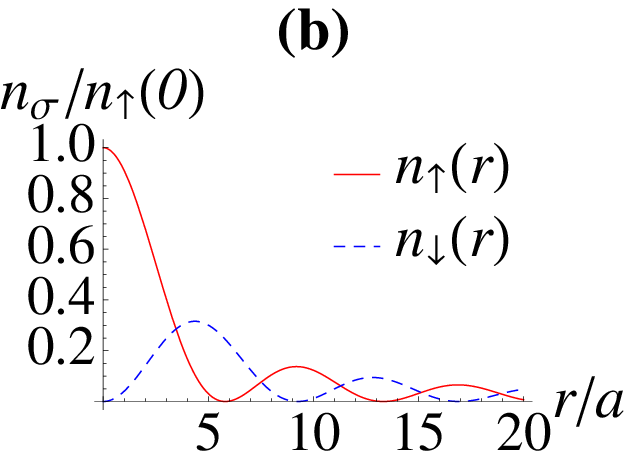}} 
    \end{tabular}
\end{center}
\caption{\label{fig:bessel1}(Color online) The radial dependence of the spin-resolved zero-energy LDOS's with 
two calculation parameter sets of 
(a) $\mu = 0.4$eV and $\Delta_{0} = 0.3$eV, and (b) $\mu = 0.5$eV and $\Delta_{0} = 0.01$eV. $a$ is the lattice constant. The spin-resolved LDOS's do not depend on $k_{z}$. }
\end{figure}

Then, one can confirm that the quasi-particle annihilation operators with the zero energy $\gamma_{\up}$ and $\gamma_{\down}$ satisfy 
the Majorana condition $\gamma_{\sigma} = \gamma_{\sigma}^{\dagger}$ because of the relation $u_{c}(\Vec{r}) = i \gamma^{2} u^{\ast}(\Vec{r})$ (see the Appendix B for more details). 
It is found that these two solutions are localized when $M_{0}^{2}<   \mu^{2} + |\Delta_{0}|^{2}$\cite{PhysRevD.81.074004}. 
Then, we can obtain the solution at finite energy by a perturbation in terms of $k_{z}$. 
The perturbed solution is expressed by the linear combination as $\psi({\bm r})^{T} = c_{\up} (u_{\up}({\bm r}),  u_{\rm c \up}({\bm r}))^{T}+ c_{\down}(u_{\down}({\bm r}),  u_{\rm c \down}({\bm r}))^{T}$.
Substituting this solution into Eq.~(\ref{eq:bdgdirac}), the energy dispersion relations are simply given by 
\begin{align}
E = \pm v k_{z},
\end{align}
where the coefficients $(c_{\up},c_{\down}) = (1,\pm i)/\sqrt{2}$ are not dependent on $k_{z}$, and  
\begin{align}
v &\equiv \frac{4 \pi \sqrt{\mu^{2} - M_{0}^{2}}}{\sqrt{\lambda_{+} \lambda_{-}}}
\int_{0}^{\infty} dr r e^{- 2 K(r)} \left( 
J_{0}(\bar{r})^{2} - J_{1}(\bar{r})^{2}
\right).
\end{align}
Then, the spin-resolved local densities of states (LDOS's) 
$n_{\up}(E =  v k_{z},r) (\equiv |\psi_{1}(r)|^{2} + |\psi_{3}(r)|^{2})$ and $n_{\down}(E =  v k_{z},r) (\equiv |\psi_{2}(r)|^{2} + |\psi_{4}(r)|^{2})$ are, respectively,  expressed as 
\begin{align}
n_{\up}(E =v k_{z},r) &\equiv \left( \frac{\mu + M_{0} }{2 \lambda_{+}} +  \frac{\mu - M_{0} }{2 \lambda_{-}} \right) J_{0}(\bar{r})^{2} e^{- 2 K(r)}, \label{eq:nup} \\
n_{\down}(E = v k_{z},r) &\equiv \left( \frac{\mu - M_{0} }{2 \lambda_{+}} +  \frac{\mu + M_{0} }{2 \lambda_{-}} \right) J_{1}(\bar{r})^{2} e^{- 2 K(r)},\label{eq:ndown} 
\end{align}
where $\psi_{i}(r)$ is the $i$-th component of the Dirac spinor $\psi(\Vec{r})$. 
The integrated magnetization near a vortex core is also calculated by 
\begin{align}
M_{z}(r) &\sim \int_{-\infty}^{0}dE ( n_{\up}(E,r)-n_{\down}(E,r)), \\
&\propto  n_{\up}(E\sim 0,r)-n_{\down}(E \sim 0,r). \label{eq:mz}
\end{align}
We display $r$-dependence of $n_{\up}$ and $n_{\down}$ in two parameter sets 
with the approximated radial function of gap $f(r) = \Delta_{0} r/\sqrt{r^{2} + 1}$ in Fig.~\ref{fig:bessel1}\cite{PhysRevB.78.064513,PhysRevB.83.104523}. 
As seen in Fig.~\ref{fig:bessel1}(a), $n_{\up}$ in $\mu = 0.4$ eV and $\Delta_{0} = 0.3$ eV
shows the peak structure at the vortex center ($r = 0$) in contrast to the 
zero density in $n_{\down}$ 
 because of $J_{0}(0) \neq 0$ and $J_{1}(0) = 0$ in Eqs.~(\ref{eq:nup}) and (\ref{eq:ndown}). 
In Fig.~\ref{fig:bessel1}(b), $\mu = 0.5$ eV and $\Delta_{0} = 0.01$ eV, which are the same as those in the previous papers \cite{PhysRevB.86.094507,PhysRevLett.107.217001}, $n_{\up}$ and $n_{\down}$ qualitatively behave in  
similar distribution patterns 
%
as shown in Fig.~\ref{fig:bessel1}(a). 

From the results [$n_{\up}(r = 0) > 0$ and $n_{\down}(r=0)=0$] and Eq.~(\ref{eq:mz}), it is found that $M_{z}(r = 0) > 0$, {\it i.e.} 
the vortex core is spin-polarized. 
When the gap function is the pseudo-scalar, the spin-polarized vortex {\it always} emerges because of $n_{\down}(r = 0) = 0$. 
We note that the local spin imbalance is always satisfied whenever the self-consistent calculation chooses the pseudo-scalar gap function, 
since $n_{\down}(r = 0) = 0$ is guaranteed by the mathematical constraint on the vortex solution.
Though the self-consistent calculation way slightly affects the gap function\cite{PhysRevB.43.7609,PhysRevLett.65.1820}, 
we confirm that the shape of the gap function just changes the intensity of the spin-polarization as discussed later.

\section{Numerical results}
\begin{figure}
\begin{center}
     \begin{tabular}{p{1 \columnwidth}} 
      \resizebox{1.03 \columnwidth}{!}{ \includegraphics{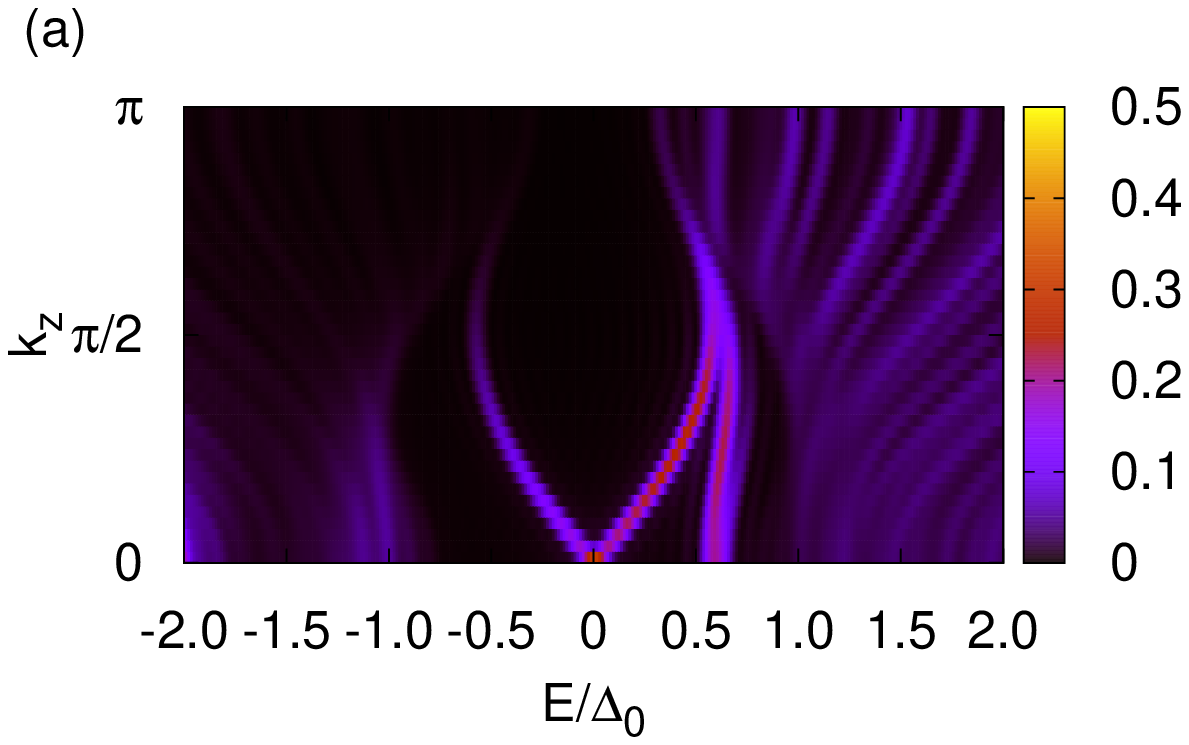}} \\
      \resizebox{1 \columnwidth}{!}{\includegraphics{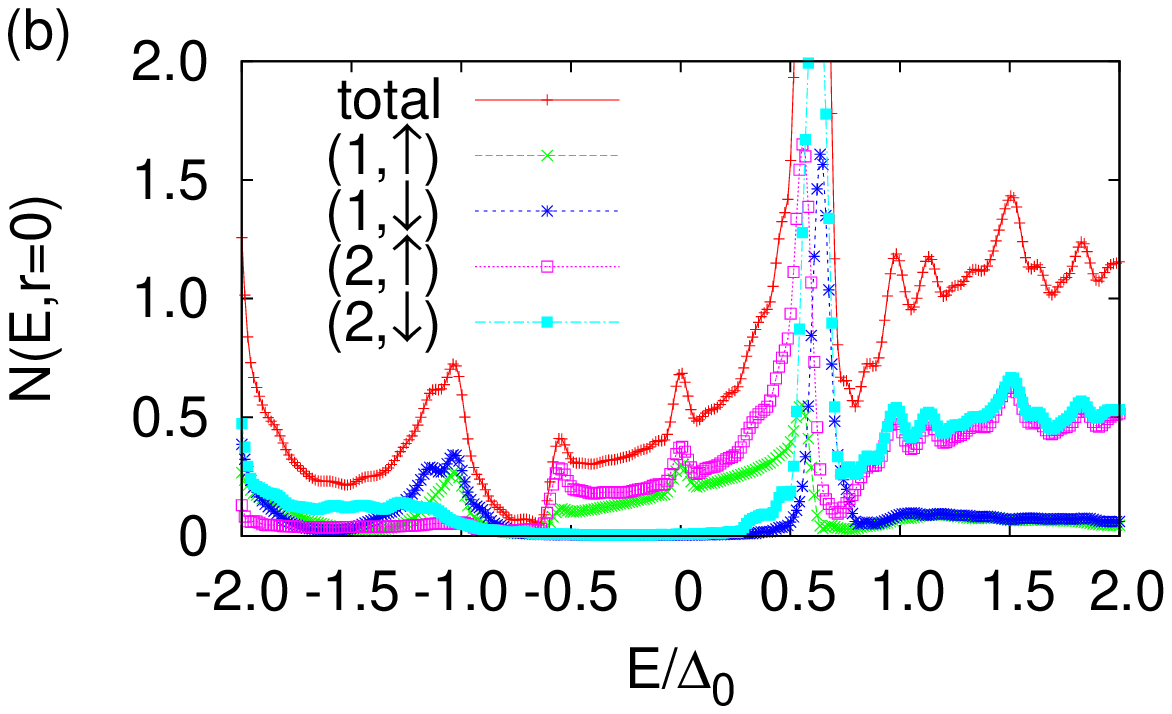}} 
    \end{tabular}
\end{center}
\caption{\label{fig:vortex}(Color online) The energy dependence of the LDOS's at the vortex center $(i_{x},i_{y}) = (L_{x}/2,L_{y}/2)$ in the pseudo-scalar superconductor. 
(a) the $k_{z}$-resolved LDOS's. (b) the total and partial LDOS's. The total peak reaches the value 4.5 (not shown). The label $(l,\sigma)$ denotes the $\sigma$-spin component with the orbital $l$.}
\end{figure}

\begin{figure}
\begin{center}
\resizebox{0.9 \columnwidth}{!}{\includegraphics{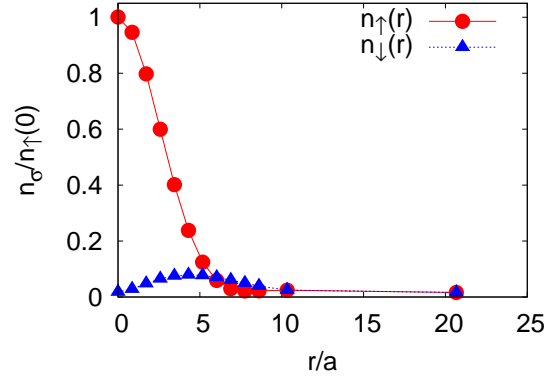}}
\end{center}
\caption{\label{fig:ldos}(Color online) The $r$-dependence of the spin-resolved LDOS's at the zero-energy in the radial direction from a vortex core. $r$ is rescaled as  
$r/a = \sqrt{3}(i_{x}-L_{x}/2) /(2 \bar{A}_{2} a)$. $i_{y} = L_{y}/2$ with $\bar{A}_{2} a = 4.1$. 
}
\end{figure}

Next, we numerically calculate the LDOS's with the use of the material parameter sets for Cu$_{x}$Bi$_{2}$Se$_{3}$ to compare with the above analytical results.
The 
$L_{x} \times L_{y}$ triangle lattice grid is employed, and  
a single vortex center is located at $(i_{x},i_{y}) = (L_{x}/2,L_{y}/2)$. 
To obtain the LDOS $n(\omega,i_{x},i_{y})$, we use the spectral polynomial expansion scheme\cite{PhysRevLett.105.167006,PhysRevB.85.092505,JPSJ.81.024710,RevModPhys.78.275} with 
 40 $k_{z}$-points and $L_{x} = L_{y} = 96$.
We take $a = 30$ eV and $b = -\mu$ as the renormalization factors, $\eta = 1 \times 10^{-3}$ eV as a smearing factor, and 
$n_{c} = 8000$ as a cut-off parameter (see, Ref.~\citen{JPSJ.81.024710}).
The gap-amplitude, $\Delta_{0} = 0.3$ eV, the chemical potential, $\mu = 0.4$ eV, 
which are the same as  the parameter set of the analytical result in Fig.~\ref{fig:bessel1}(a), and other 
parameters are the same as those in Ref.~\citen{PhysRevB.86.094507} (see the Appendix C). 
In the presence of a vortex, 
one finds the zero-energy bound states at the vortex center, in which there are 
$k_{z}$-dispersive energy spectra of two bound states as shown in Fig.~\ref{fig:vortex}(a). 
We confirm that a linear dispersion relation develops around the zero-energy while  
the other has a flat dispersion at the mid-gap energy ($E/\Delta_{0} \sim 0.5$). 
Moreover, it is found that their $k_{z}$-dispersive spectra consist of only up-spin quasiparticles 
while the mid-gap spectrum is opposite (see, Fig.~\ref{fig:vortex}(b)).  
As shown in Fig.~\ref{fig:ldos}, $r$-dependence of the spin-resolved LDOS at the zero-energy around a vortex core 
reveals that the core is spin-polarized being consistent with our analytical calculation shown in Fig.~\ref{fig:bessel1}(a).

\section{Discussion}
\subsection{Spin-polarization around a vortex}
Finally, we discuss the reason why the Majorana fermion is spin-polarized. 
It is well known that the state characterized by the zero orbital angular momentum ($L_{z} = 0$) has the minimum absolute energy 
with a given $k_{z}$ around a vortex. 
On the other hand, 
the {\it total} angular momentum ${\bm J}$ substitutes for ${\bm L}$ in the present system according to the spin-orbit coupling. 
Thus, the bound state with the minimum absolute value of $J_{z}$ has the minimum absolute energy. 
The possible total angular momentum is $J_{z} = 1/2$ or $-1/2$, and which of them is selected as the lowest energy bound state depends on the vortex line direction. 
It should be noted that the orbital and spin angular momenta ${\bm L}$ and ${\bm S}$ are 
not good quantum numbers in this system. 
Thus, the eigenstate with the minimum $J_{z} = 1/2$ are given as a linear combination of 
the two states with $(L_{z},S_{z}) = (0,1/2)$ and $(1,-1/2)$. 
This indicates that the ($L_{z} = 0$) states including finite $k_{z}$-component coincide with $S_{z} = \frac{1}{2}$. 
The vortex center is clearly found to be spin-polarized, since the zero angular momentum states 
have more contribution to the wave function weight at the vortex center ($r = 0$) than non-zero angular momentum states (i.e., $J_{0}(r = 0) > J_{1}(r = 0)$).  
Thus, we conclude that  the zero-energy Majorana fermion is distinctly spin-polarized in the vortex core and 
the core itself is also spin-polarized $(M_{z}(r=0)>0)$. 
The spin-polarized core is easily observable if a vortex lattice is formed.
For example, the neutron or muon scattering is sensitive to such 
spin-density lattice modulations, and NMR is also a good probe.

A spin-polarized core never occur in the even parity scalar-type superconductivity  (so-called $\Delta_{1}$ shown as 
\(
\Delta^{11}_{\uparrow\downarrow}
=
- \Delta^{11}_{\downarrow\uparrow}
\), 
\(
\Delta^{22}_{\uparrow\downarrow}
=
- \Delta^{22}_{\downarrow\uparrow}
\), 
\(
\Delta^{11}_{\uparrow\downarrow}
=
\Delta^{22}_{\uparrow\downarrow}
\)
in Ref.~\citen{PhysRevLett.109.217004}
), since a bound state has finite energy and consists of the Bessel function with half integers ({\it e.g.} $J_{n\pm1/2}(r)$). 
In other odd-parity superconductors with polar-vector-type gap functions, 
spin-polarized Majorana bound states occur in magnetic fields parallel to a direction of the point-nodes.

Here, we discuss whether a spin-polarization can occur even for nonzero energy bound states in a vortex core which are 
not Majorana bound states. 
There are two kinds of finite energy bound states. 
One is a bound state with a zero angular momentum and finite $k_{z}$, whose spin-resolved LDOS is expressed in Eqs.~(\ref{eq:nup}) and (\ref{eq:ndown}). 
This bound state is spin-polarized and a integrated magnetization Eq.~(\ref{eq:mz}) is finite. 
The other is a bound state with a finite angular momentum, which can not be obtained analytically.  
This state consists of the Bessel functions with finite integers, i.e.  $J_{n}(r)$ and $J_{n+1}(r)$. 
A spin imbalance of a finite angular momentum state might be smaller than that of a zero angular momentum state at a vortex center because of $J_{n}(r = 0) = J_{n+1}(r=0) = 0$.

\subsection{Robustness of the spin-polarized Majorana bound states}
\begin{figure}[tb]
\begin{center}
      \resizebox{0.8 \columnwidth}{!}{\includegraphics{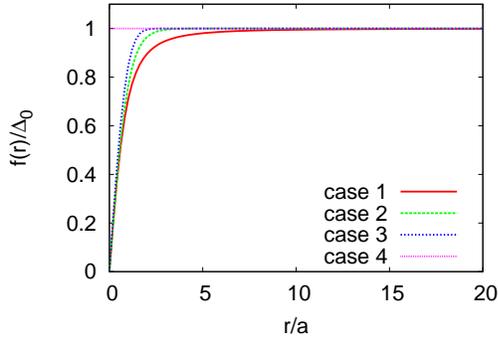}} 
\end{center}
\caption{\label{fig:kansu}(Color online) 
Various kinds of the functions $f(r)$ determined in Eq.~(\ref{eq:fr}).}
\end{figure}

\begin{figure*}[bht]
\begin{center}
     \begin{tabular}{p{0.5\columnwidth} p{0.5 \columnwidth} p{0.5 \columnwidth}p{0.5 \columnwidth}}
      \resizebox{0.5 \columnwidth}{!}{\includegraphics{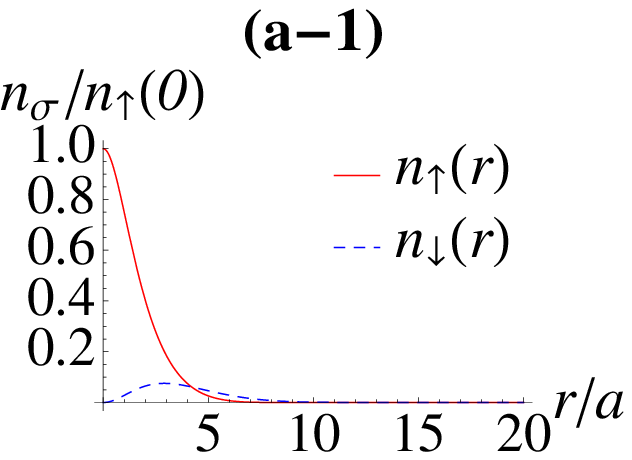}} &
      \resizebox{0.5 \columnwidth}{!}{\includegraphics{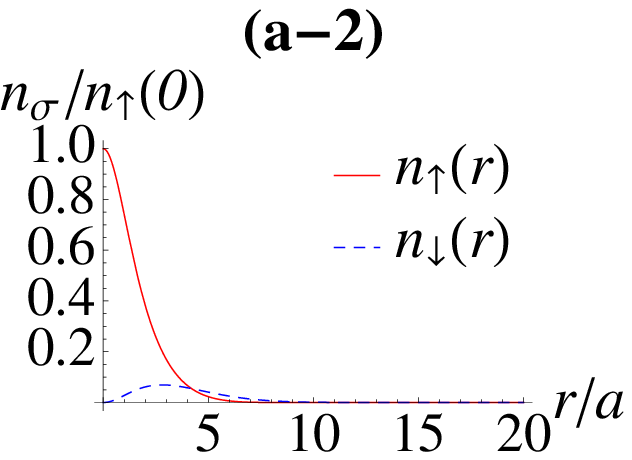}} &
      \resizebox{0.5 \columnwidth}{!}{\includegraphics{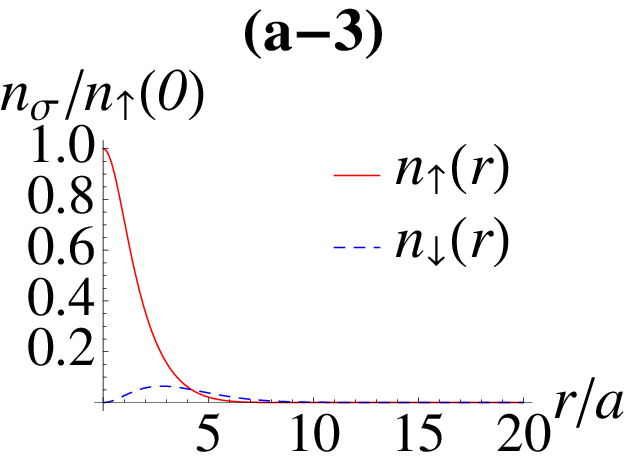}} &
      \resizebox{0.5 \columnwidth}{!}{\includegraphics{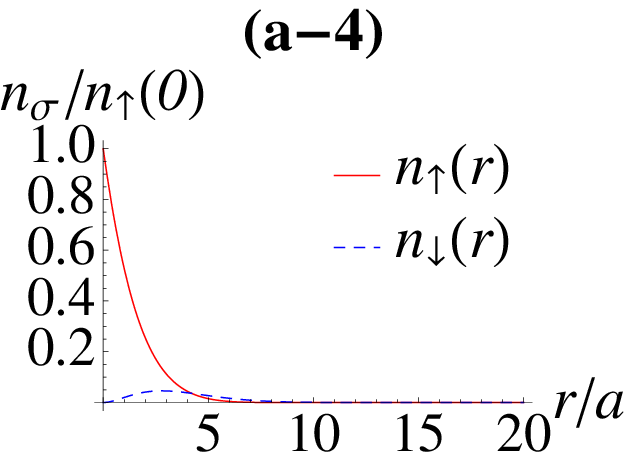}}  \\
       \resizebox{0.5 \columnwidth}{!}{\includegraphics{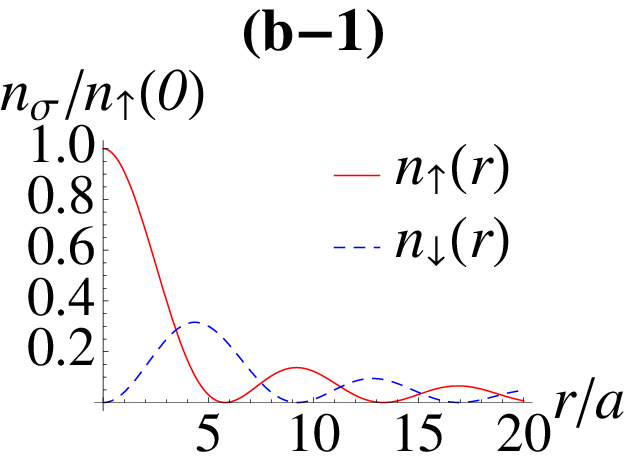}} &
      \resizebox{0.5 \columnwidth}{!}{\includegraphics{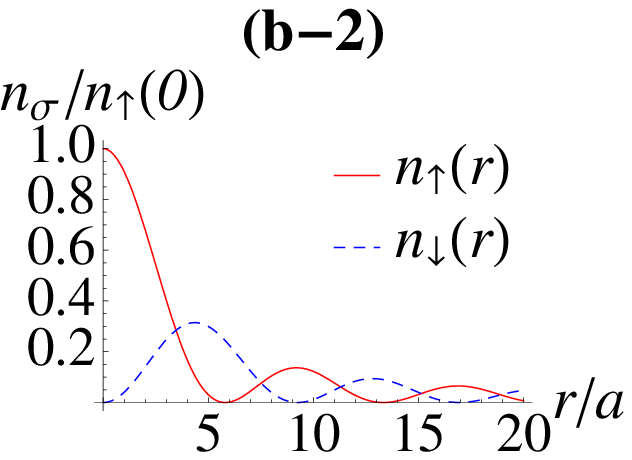}} &
      \resizebox{0.5 \columnwidth}{!}{\includegraphics{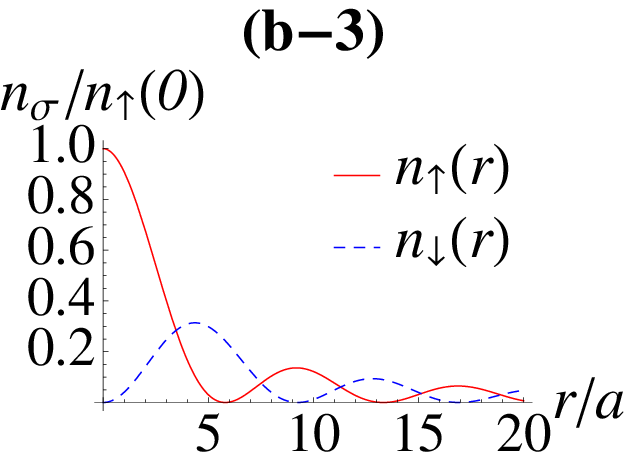}} &
      \resizebox{0.5 \columnwidth}{!}{\includegraphics{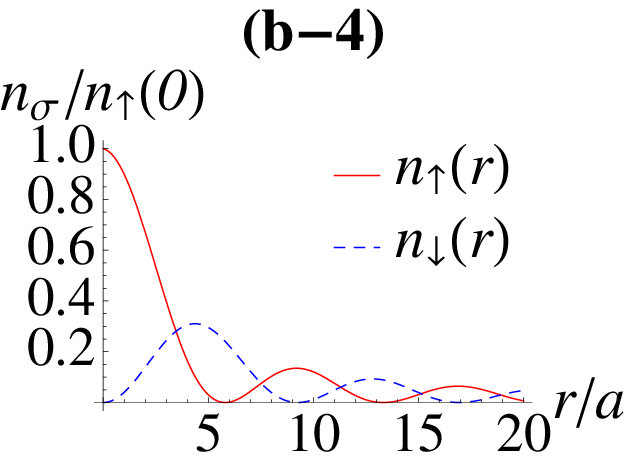}} 
    \end{tabular}
\end{center}
\caption{\label{fig:fr}(Color online) The radial dependence of the spin-resolved zero-energy LDOS's with 
two calculation parameter sets of 
(a) $\mu = 0.4$eV and $\Delta_{0} = 0.3$eV, and (b) $\mu = 0.5$eV and $\Delta_{0} = 0.01$eV. $a$ is the lattice constant. 
The function $f(r)$ in Eq.~(\ref{eq:fr}) is adopted, respectively.}
\end{figure*}

We show the robustness of the spin-polarized Majorana bound states. 
In general, solving the gap equations closes the self-consistent calculations in the Bogoliubov-de Gennes formalism. 
We adopt the approximated radial function of gap $f(r) = \Delta_{0} r/\sqrt{r^{2}+1}$ in the previous section. 
The self-consistent calculation yields the correct form of $f(r)$. 
We have to show that the self-consistent calculation does not change our results. 
However, one can not determine the parameters in the gap equations, 
since the kind of the pairing interaction for Cu$_{x}$Bi$_{2}$Se$_{3}$ has not been determined in experiments. 
Therefore, we show the $f(r)$-dependence of the spin-polarized bound states in this section. 
The correct form of the function $f(r)$ must satisfy the condition of $f(r = 0) = 0$ and $\lim_{r \rightarrow \infty} f(r) = \Delta_{0}$.
We consider the several functions as follows (See, Fig.~\ref{fig:kansu}):
\begin{align}
f(r) &= 
\left\{ \begin{array}{ll}
\Delta_{0} {\displaystyle \frac{r}{\sqrt{r^{2}+ 1}}} & ({\rm case \: 1}) \\
\Delta_{0} \tanh(r) & ({\rm case \: 2}) \\
\Delta_{0} {\rm Erf}(r) & ({\rm case \: 3}) \\
\Delta_{0}  & ({\rm case \: 4}) 
\end{array}. \right. \label{eq:fr}
\end{align}
As shown in Fig.~\ref{fig:fr}, we can conclude that the results do not depend on the form of $f(r)$  in terms of the spin-polarization. 

\section{Conclusion}
In conclusion, we examined vortex core bound-states in the topological superconductors. 
With the use of the analytical and numerical calculations, we found that the zero-energy Majorana fermions are spin-polarized and the vortex core itself is distinctly magnetized when the rotational symmetry around the vortex line is preserved. 
The result is universal for Dirac superconductivity whose rotational degree of freedom is characterized by 
total angular momentum $J = S + L$. 
Such a drastic feature is easily observable in experiments.

\section*{Acknowledgment}
We thank M. Okumura for helpful discussions and comments. The calculations were performed using the supercomputing 
system PRIMERGY BX900 at the Japan Atomic Energy Agency. 
This study was supported by Grants-in-Aid for Scientific Research from MEXT of Japan.

\onecolumn
\appendix
\section{Solutions of the BdG equation around a vortex}
Let us solve the following BdG equations:
\begin{align}
\hat{H}
\left(\begin{array}{c}
u({\bm r}) \\
u_{\rm c}({\bm r})
\end{array}\right)
&= E \left(\begin{array}{c}
u({\bm r}) \\
u_{\rm c}({\bm r})
\end{array}\right), 
\end{align}
where the Hamiltonian is 
\begin{align}
\hat{H} &= 
\left(\begin{array}{cc} \gamma^{0} \hat{H}^{-}({\bm r})  &\gamma^{0} \Delta^{-}({\bm r})\\
\gamma^{0} \Delta^{+}({\bm r}) & \gamma^{0} \hat{H}^{+}({\bm r}) \end{array}\right)
= 
\left(\begin{array}{cc} 
\hat{h}({\bm r}) -\mu &
\Delta'({\bm r})\\
\Delta^{'\dagger} &  \hat{h}({\bm r}) + \mu
 \end{array}\right),
\end{align}
with 
\begin{align}
\hat{h}({\bm r}) &= \gamma^{0} M_{0} - i \partial_{x} \gamma^{0} \gamma^{1} - i \partial_{y} \gamma^{0} \gamma^{2} 
- i \partial_{z} \gamma^{0} \gamma^{3}.
\end{align}
It should be noted that one can find the solution $(u^{-},u_{c}^{-}) = (i \gamma^{2} u_{c}^{\ast}, i \gamma^{2} u^{\ast})$ with 
the energy $-E$ when $(u,u_{c})$ is a solution with the energy $E$. 
The zero-energy solutions satisfy the following equations, 
\begin{align}
(\hat{h}-\mu) u + \Delta' i \gamma^{2} u^{\ast} &= 0, \label{eq:dif}\\
\Delta^{' \dagger} i \gamma^{2} u_{c}^{\ast} + (\hat{h} + \mu) u_{c} &= 0.
\end{align}
We assume that these solutions are expressed as 
\begin{align}
u(\Vec{r}) &= \zeta(r) u^{\rm N}(r,\theta,z). \label{eq:zeta}
\end{align}
Here, $u^{\rm N}(r,\theta,z)$ is a solution in the normal states which satisfies
\begin{align}
(\hat{h}-\mu) u^{\rm N}(\Vec{r}) &= 0.
\end{align}

At first, we solve the equation in the normal states. 
The equation is rewritten as 
\begin{align}
\hat{h}^{2} u^{\rm N}(\Vec{r}) &= \mu^{2} u^{\rm N}(\Vec{r}), \\
(-\Vec{\nabla}^{2} + M_{0}^{2})u^{\rm N}(\Vec{r}) &= \mu^{2} u^{\rm N}(\Vec{r}).
\end{align}
Thus, each component of the solution is expressed as 
\begin{align}
u^{{\rm N} i}(\Vec{r}) &= c_{i} e^{i n_{i} \theta} J_{n_{i}}(\alpha  r) e^{i k_{z} z},
\end{align}
with $\alpha \equiv \sqrt{\mu^{2} - M_{0}^{2} - k_{z}^{2}}$. 
In order to obtain coefficients $(c_{1},c_{2},c_{3},c_{4})$, we solve the following equations,
\begin{align}
\left(\begin{array}{cccc}
M_{0} & 0 & k_{z} & L_{-} \\
0 & M_{0} & L_{+} & -k_{z} \\
k_{z} & L_{-} & -M_{0} & 0 \\
L_{+} & -k_{z} & 0 & -M_{0}
\end{array}\right)
\left(\begin{array}{c}
c_{1} e^{i n_{1} \theta} J_{n_{1}}(\alpha r) \\
c_{2} e^{i n_{2} \theta} J_{n_{2}}(\alpha r) \\
c_{3} e^{i n_{3} \theta} J_{n_{3}}(\alpha r) \\
c_{4} e^{i n_{4} \theta} J_{n_{4}}(\alpha r)
\end{array}\right)e^{i k_{z} z} &=
\mu \left(\begin{array}{c}
c_{1} e^{i n_{1} \theta} J_{n_{1}}(\alpha r) \\
c_{2} e^{i n_{2} \theta} J_{n_{2}}(\alpha r) \\
c_{3} e^{i n_{3} \theta} J_{n_{3}}(\alpha r) \\
c_{4} e^{i n_{4} \theta} J_{n_{4}}(\alpha r)
\end{array}\right)e^{i k_{z} z},
\end{align}
with 
\begin{align}
L_{\pm} &\equiv \pm e^{\pm i \theta} (\mp i \frac{\partial}{\partial r} + \frac{1}{r} \frac{\partial}{\partial \theta} ). 
\end{align}
With the use of the relations $\partial_{r} J_{n}(r) = - J_{n+1}(r) + (n/r) J_{n}(r)$ and 
$\partial_{r} J_{n}(r) =  J_{n-1}(r) - (n/r) J_{n}(r)$
, we obtain 
\begin{align}
L_{\pm} e^{i n \theta} J_{n}(\alpha r) &= \pm e^{i (n \pm 1) \theta} (\mp i \frac{\partial}{\partial r} + \frac{ i n}{r} )J_{n}(\alpha r) \\
&=  \pm e^{i (n \pm 1) \theta}
 i \alpha J_{n\pm1}(\alpha r).
\end{align}
Thus, the equations are rewritten as 
\begin{align}
\left(\begin{array}{cccc}
(M_{0}-\mu)e^{i n_{1} \theta} J_{n_{1}}(\alpha r) & 0 & k_{z}e^{i n_{3} \theta} J_{n_{3}}(\alpha r) &
- e^{i (n_{4} - 1) \theta}
 i \alpha J_{n_{4}-1}(\alpha r)
 \\
0 & (M_{0}-\mu)e^{i n_{2} \theta} J_{n_{2}}(\alpha r)  &
 e^{i (n_{3} + 1) \theta}
 i \alpha J_{n_{3}+1}(\alpha r)
  & -k_{z}e^{i n_{4} \theta} J_{n_{4}}(\alpha r)  \\
k_{z}e^{i n_{1} \theta} J_{n_{1}}(\alpha r)  & 
- e^{i (n_{2} - 1) \theta}
 i \alpha J_{n_{2}-1}(\alpha r)
 & (-M_{0}-\mu)e^{i n_{3} \theta} J_{n_{3}}(\alpha r)  & 0 \\
 e^{i (n_{1} + 1) \theta}
 i \alpha J_{n_{1}+1}(\alpha r)
& -k_{z} e^{i n_{2} \theta} J_{n_{2}}(\alpha r) 
 & 0 &
 (-M_{0}-\mu)e^{i n_{4} \theta} J_{n_{4}}(\alpha r) 
\end{array}\right)
\left(\begin{array}{c}
c_{1}  \\
c_{2} \\
c_{3}  \\
c_{4} 
\end{array}\right) &= 
0.
\end{align}
The orbital angular momenta $n_{i}$ must satisfy
\begin{align}
n_{2} &= n_{1} + 1, \: \: n_{3} = n_{1}, \: \: n_{4} = n_{1} + 1.
\end{align}
Then, in the case of $k_{z} = 0$, 
the coefficients are written as 
\begin{align}
c_{3} &= -\frac{i c_{2} \alpha}{M_{0}+\mu}, \: \: c_{4} = \frac{i c_{1} \alpha}{M_{0}+\mu}.
\end{align}
We assume that $\zeta(r)$ in Eq.~(\ref{eq:zeta}) is a real scalar function. 
By substituting Eq.~(\ref{eq:zeta}) into the differential equations Eq.~(\ref{eq:dif}), we obtain 
\begin{align}
\left(\begin{array}{c}
 i e^{- i \theta}   \frac{\partial \zeta(r)}{\partial r}\frac{i c_{1} \alpha}{M_{0}+\mu} e^{i (n_{1}+1) \theta} J_{n_{1}+1}(\alpha r)\\
- i e^{ i \theta}   \frac{\partial \zeta(r)}{\partial r}\frac{i c_{2} \alpha}{M_{0}+\mu} e^{i n_{1} \theta} J_{n_{1}}(\alpha r) \\
 i e^{- i \theta}   \frac{\partial \zeta(r)}{\partial r}c_{2} e^{i (n_{1}+1) \theta} J_{n_{1}+1}(\alpha r)\\
i e^{ i \theta}   \frac{\partial \zeta(r)}{\partial r}c_{1} e^{i n_{1} \theta} J_{n_{1}}(\alpha r)
\end{array}\right)
+ \zeta(r) \Delta^{'}(r)
\left(\begin{array}{c}
\frac{-i c_{1}^{\ast} \alpha}{M_{0}+\mu} e^{-i (n_{1}+1) \theta} J_{n_{1}+1}(\alpha r)\\
\frac{i c_{2}^{\ast} \alpha}{M_{0}+\mu} e^{-i n_{1} \theta} J_{n_{1}}(\alpha r) \\
-c_{2}^{\ast} e^{-i (n_{1}+1) \theta} J_{n_{1}+1}(\alpha r) \\
-c_{1}^{\ast} e^{-i n_{1} \theta} J_{n_{1}}(\alpha r)
\end{array}\right)
 &= 0
\end{align}

Finally, we consider the gap function represented by s pseudo-scalar.
In this case, $\Delta'(\Vec{r})$ with a vortex is expressed as 
\begin{align}
\Delta'(\Vec{r}) &= f(r) e^{i M \theta} \gamma^{0}.
\end{align}
Here, $M$ is a winding number and $f(r)$ is the amplitude of the order parameters ($f(r=0) = 0$ and $f(r) > 0$).
By substituting $\Delta'(r)$ into the above differential equations, we obtain 
\begin{align}
\left(\begin{array}{c}
 i   \frac{\partial \zeta(r)}{\partial r}\frac{i c_{1} \alpha}{M_{0}+\mu} e^{i n_{1} \theta} J_{n_{1}+1}(\alpha r)\\
- i   \frac{\partial \zeta(r)}{\partial r}\frac{i c_{2} \alpha}{M_{0}+\mu} e^{i (n_{1}+1) \theta} J_{n_{1}}(\alpha r) \\
 i   \frac{\partial \zeta(r)}{\partial r}c_{2} e^{i n_{1} \theta} J_{n_{1}+1}(\alpha r)\\
i   \frac{\partial \zeta(r)}{\partial r}c_{1} e^{i (n_{1}+1) \theta} J_{n_{1}}(\alpha r)
\end{array}\right)
+ \zeta(r) f(r)
\left(\begin{array}{c}
\frac{-i c_{1}^{\ast} \alpha}{M_{0}+\mu} e^{i (-n_{1}-1+M) \theta} J_{n_{1}+1}(\alpha r)\\
\frac{i c_{2}^{\ast} \alpha}{M_{0}+\mu} e^{i (-n_{1}+M) \theta} J_{n_{1}}(\alpha r) \\
-c_{2}^{\ast} e^{i (-n_{1}-1+M) \theta} J_{n_{1}+1}(\alpha r) \\
-c_{1}^{\ast} e^{i (-n_{1}+M) \theta} J_{n_{1}}(\alpha r)
\end{array}\right)
 &= 0.
\end{align}
The angular momentum of the zero energy solutions is 
\begin{align}
n_{1} &= \frac{M-1}{2}.
\end{align}
Since $\zeta(r)$ is a scalar real function, 
we obtain two solutions with coefficients
\begin{align}
(c_{1},c_{2}) &= (|c_{1}| \exp \left[ \frac{\pi i }{4} \right],0), \: \:  (0,|c_{2}| \exp \left[ -\frac{\pi i }{4} \right]).
\end{align}
Therefore, the zero-energy solutions with a winding number $M$ are written as 
\begin{align}
u_{\up}(\Vec{r}) &= c_{\up} e^{-K(r)} 
\left(\begin{array}{c}
b_{+} \exp \left[ i \frac{M-1}{2} \theta \right] J_{\frac{M-1}{2}}(a r) \\
0 \\
0 \\
i  b_{-}
 \exp \left[ i \frac{M+1}{2} \theta \right] J_{\frac{M+1}{2}}(a r)
\end{array}\right), \\
u_{\down}(\Vec{r}) &= c_{\down} e^{-K(r)} 
\left(\begin{array}{c}
0 \\
b_{+} \exp \left[ i \frac{M+1}{2} \theta \right] J_{\frac{M+1}{2}}(a r) \\
-i  b_{-} \exp \left[ i \frac{M-1}{2} \theta \right]  J_{\frac{M-1}{2}}(a r) \\
0
\end{array}\right),
\end{align}
with $a = \sqrt{\mu^{2} - M_{0}^{2}}$, $b_{\pm} = \sqrt{\mu \pm M_{0}}$ and $K(r) = \int_{0}^{r} |f(r')| dr'$.
\section{Proof of the Majorana condition}
In terms of the BdG equations, the quasiparticle annihilation operator $\gamma_{\sigma}$ with the zero energy is expressed as 
\begin{align}
\gamma_{\sigma} &= \int d\Vec{r}  \left\{ u_{\sigma}^{\rm T}(\Vec{r}) \psi(\Vec{r}) + u_{{\rm c},\sigma}^{\rm T} \psi_{\rm c}(\Vec{r}) \right\}. \label{eq:ani}
\end{align}
Here, $\psi_{\rm c} = i \gamma^{2} (\psi^{\dagger})^{\rm T}$. At the zero energy, there is a relation expressed as $u_{{\rm c},\sigma}(\Vec{r}) = i \gamma^{2} u_{\sigma}^{\ast}(\Vec{r})$.
Substituting the above relation into eq.~(\ref{eq:ani}), the quasiparticle creation operator $\gamma_{\sigma}^{\dagger}$ 
is written as 
\begin{align}
\gamma_{\sigma}^{\dagger} &= \int d\Vec{r} \left\{  \psi(\Vec{r})^{\dagger} u_{\sigma}^{\ast}(\Vec{r})  + \right[ \psi^{\rm T}(\Vec{r}) (i \gamma^{2})^{\dagger} \left] \left[ i \gamma^{2} \right]^{\ast} u_{\sigma}^{\ast}(\Vec{r})  \right\} , \\
&= \gamma_{\sigma}.
\end{align}
Thus, the quasiparticle creation operator $\gamma_{\sigma}^{\dagger}$ 
with the zero energy eigenvalue satisfies the Majorana condition. 

\twocolumn
\section{Parameters in numerical calculations}
We show the parameters in numerical calculations.
The mean-field Hamiltonian on the triangular lattice based on the Bogoliubov-de Gennes formalism is 
expressed as 
\begin{align}
H = \sum_{k_{z}} \sum_{i,j} 
\left(\begin{array}{cc}c_{i}^{\dagger} & c_{i}^{T}
\end{array}\right)
\left(\begin{array}{cc}\hat{H}_{ij}(k_{z}) & \hat{\Delta} f({\bm R}_{i}) \delta_{ij} \\
\hat{\Delta}^{\dagger} \delta_{ij} f({\bm R}_{i})^{\ast}& - \hat{H}_{ij}^{\ast}(-k_{z})
\end{array}\right)
\left(\begin{array}{c}c_{j}  \\c_{j}^{\ast}
\end{array}\right),
\end{align}
where $c_{i}^{\dagger}$ is the $4$-component creation operator at the $i$-th site on the two-dimensional triangle lattice and 
$k_{z}$ denotes the momentum in the crystal $c$-axis. 
$\hat{\Delta} = \gamma^{0} \Delta^{-} i \gamma^{2} \gamma^{0}$ is $4 \times 4$ matrix whose elements are given as $\Delta_{\sigma \sigma'}^{lm}$ with orbital $l(m)$ and spin $\sigma(\sigma')$ indices. 
The normal state Hamiltonian $\hat{H}_{ij}(k_{z})$ is given by 
\begin{align}
 \hat{H}_{ij}(k_{z})  &=  \int d{\bm k}_{\perp} e^{i {\bm k}_{\perp} \cdot ({\bm R}_{i} - {\bm R}_{j})} \hat{H}({\bm k}_{\perp},k_{z}),
\end{align}
with $ab$-plane momentum ${\bm k}_{\perp} = (k_{x},k_{y})$. 
The $4 \times 4$ matrix $\hat{H}({\bm k}_{\perp},k_{z})$ is expressed as 
\begin{align}
\hat{H}({\bm k}_{\perp},k_{z}) &= M({\bm k}_{\perp},k_{z}) \gamma^{0} + P_{0}({\bm k}_{\perp},k_{z}) \nonumber \\
&+ \gamma^{0} P_{1}({\bm k}_{\perp}) \gamma^{1}
+ \gamma^{0} P_{2}({\bm k}_{\perp}) \gamma^{2} + \gamma^{0} P_{3}(k_{z}) \gamma^{3} ,
\end{align}
where, 
\begin{align}
M({\bm k}_{\perp},k_{z}) &\equiv M_{0} - 2 \bar{B}_{1}(1 - \cos (k_{z}))- \bar{B}_{2}\eta({\bm k}_{\perp}), 
\end{align}
\begin{align}
P_{0}({\bm k}_{\perp},k_{z}) &\equiv 2 \bar{D}_{1}(1 - \cos (k_{z})) + \bar{D}_{2} \eta({\bm k}_{\perp}) - \mu, \\
P_{1}({\bm k}_{\perp}) &\equiv \frac{2}{3} \bar{A}_{2} \sqrt{3} \sin \left( \frac{\sqrt{3}}{2} k_{x} \right) \cos \left(\frac{k_{y}}{2} \right), \\
P_{2}({\bm k}_{\perp}) &\equiv \frac{2}{3}  \bar{A}_{2} \left(\cos \left( \frac{\sqrt{3}}{2} k_{x} \right) \sin \left(\frac{k_{y}}{2} \right) + \sin(k_{y})\right), \\
P_{3}(k_{z}) &\equiv \bar{A}_{1} \sin(k_{z}), 
\end{align}
with 
$\eta({\bm k}_{\perp}) \equiv (3 - 2 \cos(\sqrt{3} k_{x}/2)\cos(k_{y}/2) - \cos(k_{y}))$.
We set 
$M_{0} = 0.28$ eV, $\bar{A}_{1} = 0.32$ eV, $\bar{A}_{2} = 4.1/a$ eV, $\bar{B}_{1} = 0.216$ eV, $\bar{B}_{2} = 56.6/a^{2}$ eV, $\bar{D}_{1} = 0.024$ eV, $\bar{D}_{2} = 19.6/a^{2}$ and $a =  4.076$ \AA  \: as the material parameters for Cu$_{x}$Bi$_{2}$Si$_{3}$.
For simplicity, we do not solve the gap-equation but use a spatial distribution form of the order parameter around a single vortex $f({\bm R}_{i})$
%
 written as 
\begin{align}
f({\bm R}_{i}) &= e^{i \theta} \Delta_{0} \frac{|{\bm R}_{i}|}{\sqrt{|{\bm R}_{i}|^{2} + \xi^{2}}},
\end{align}
where $\theta$ denotes the polar angle around $c$-axis, $\Delta_{0}$ is the amplitude of the order-parameter and $\xi$ is the coherence length.

\pagebreak

\end{document}